30 May 2003

# A Sum Rule for Nonlinear Optical Susceptibilities

**Luciana C. Dávila Romero and David L. Andrews**

*School of Chemical Sciences, University of East Anglia,*

*Norwich, NR4 7TJ U.K.*

**Abstract**

It is explicitly shown, for optical processes arbitrarily comprising two-, three- or four-photon interactions, that the sum over all matter states of any optical susceptibility is exactly zero. The result remains true even in frequency regions where damping is prominent. Using a quantum electrodynamical framework to render the photonic nature of the fundamental interactions, the result emerges in the form of a traceless operator in Hilbert space. The generality of the sum rule and its significance as a thermodynamic limit are discussed, and the applicability to real systems is assessed.

## 1.    Introduction

In the theory of linear and nonlinear optical response from atomic and molecular systems, a common expedient for calculational and interpretive simplicity is the adoption of a two-level approximation, in which it is assumed that any material of interest has only one significant electronic excited state in the relevant frequency range.  It has been known for some time that this approximation leads to an excited state polarisability that is the exact negative of the ground state polarisability, and the same proves true for the ground and excited state first hyperpolarisability.  In the bulk, similar relationships hold for the linear and second order optical susceptibility tensors, and the consequential possibility of observing novel harmonic resurgence phenomena in suitable media has been noted [1].  Recently, it has also been shown that the susceptibility relationships retain validity even when resonance damping is entertained [2, 3], provided the correct choice is made for the sign of the damping corrections, as exacted by the principles of time-reversal invariance.

In this paper we develop a more general theory by considering multi-level systems with an arbitrary number of excited states, also considering higher-order optical processes such as those involved in four-wave mixing.  Equations are developed using a quantum electrodynamical framework that faithfully renders the photonic nature of the fundamental interactions involved.  As a result it is explicitly shown, for two-, three- and four-photon interactions, that *for any optical process the sum of the susceptibilities for all matter states is zero*.  The result remains true even in frequency regions where damping is prominent.  Applications range from Rayleigh scattering, through harmonic generation to CARS, the ac Stark effect and phase-conjugate reflection.  The generality of the sum rule and its physical significance are discussed, and the paper concludes with an assessment of the applicability to real systems.

## 2. Molecular operators and optical susceptibilities

For each kind of electromagnetic interaction between matter and radiation – such as the absorption or emission of single photons, nonlinear optical interactions, and also many cooperative or concerted energy transfer processes, the propensity of any given molecule is quantified by a microscopic susceptibility tensor. By methods that have been extensively detailed elsewhere, general expressions for each such tensor can be derived using quantum electrodynamics (QED) formulation [4, 5]. The quantum Hamiltonian for a sub-system comprising any one optical centre $x$ and the radiation field can be separated into two terms: one is the unperturbed Hamiltonian $H_0$, which has contributions due to the radiation, $H_{rad}$, and also the Schrödinger operator for each optical centre, $H_{mol}(x)$; the second term is the corresponding electromagnetic interaction, $H_{int}(x)$. The ensemble response for any process based on one-centre interactions is then evaluated from the probability amplitude;

$$A \sim \sum_{x} \langle f | H_{int}(x) + H_{int}(x) T_{sub}(x) H_{int}(x) | i \rangle \,, \tag{1}$$

where $T_{sub}(x) \equiv \sum_{p=0}^{\infty} \left[ (E_i - H_0)^{-1} H_{int}(x) \right]^p (E_i - H_0)^{-1}$ is the resolvent operator for the sub-system, the kets $|i\rangle$ and $|f\rangle$ signifying its initial and final states, respectively. A more detailed description of this formalism can be found elsewhere [6]. For an $n$-photon process the leading contribution in the perturbative development of the probability amplitude is the matrix element $S^{fi(n)}(x)$;

$$S^{fi(n)}(x) = \langle f | H_{int}(x) \prod_{j=2}^{n} \left\{ \sum_{s^{(j)}} (E_0 - H_0)^{-1} | s^{(j)} \rangle \langle s^{(j)} | H_{int}(x) \right\} | i \rangle \bigg|_{x} \,, \tag{2}$$

where $\left| s^{(j)} \right\rangle\big|_{x}$ are intermediate sub-system states, elements of a set comprising direct products of all radiation and matter eigenstates of $H_0$. Writing explicitly these states we have;

$$\left|s^{(j)}\right\rangle_x = \left|s_j(x)\right\rangle \otimes \left|\text{rad}_j\right\rangle. \tag{3}$$

Here the first label corresponds to the material state and the second one corresponds to the radiation. In similar way the initial and final states of the sub-system can be represented as;

$$\left|f\right\rangle_x = \left|s_f(x)\right\rangle \otimes \left|\text{rad}_f\right\rangle, \tag{4}$$

$$\left|i\right\rangle_x = \left|s_i(x)\right\rangle \otimes \left|\text{rad}_i\right\rangle. \tag{5}$$

In the usual electric-dipole approximation the interaction Hamiltonian is $H_{\text{int}}(x) = -e_0^{-1} m_i(x) d_i^\perp(\mathbf{R}_x)$, coupling the electric dipole operator $\mathbf{\mu}(x)$ and the transverse part of the electric displacement field operator $\mathbf{d}(\mathbf{R}_x)$, using the implied summation convention for repeated Cartesian indices. From here onwards the label $x$ is dropped to simplify notation. In view of the usual mode expansion of $\mathbf{d}(\mathbf{R}_x)$ in terms of creation and annihilation operators, the linearity of the interaction Hamiltonian itself ensures that each operation either creates or destroys a photon. In equation (2), the sum over all states and the product over $j$ thus serve to identify all possible sequences of allowed routes between the initial and final radiation state – corresponding to the familiar time orderings or state-sequences. When the explicit form of the interaction Hamiltonian is substituted, the quantum amplitude can neatly be written as the product of two tensors [7]:

$$S^{fi(n)}(x) = \mathbf{g}_{i_1...i_n}^{\text{rad}_f;\text{rad}_i} \otimes O_{i_1...i_n}^{s_f s_i}(x), \tag{6}$$

where $O_{i_1...i_n}^{s_f s_i}(x)$ is the nonlinear susceptibility tensor, and

$$\mathbf{g}_{i_n...i_1}^{\text{rad}_f \text{rad}_i} = \left\langle \text{rad}_{ini}\left|e_0^{-1} d_{i_n}^\perp\right|\text{rad}_N\right\rangle\left\langle \text{rad}_N\left|e_0^{-1} d_{i_{n-1}}^\perp\right|\text{rad}_{N-1}\right\rangle...\left\langle \text{rad}_2\left|e_0^{-1} d_{i_1}^\perp\right|\text{rad}_i\right\rangle, \tag{7}$$

delivers a tensor comprising a product of polarization components. The former, the main interest of this work, can be expressed as

$$O_{i_1...i_n}^{s_f s_i}(\mathbf{x}) = \sum_{p[\{j\},n]} \left\langle s_f \left| \mathbf{m}_{i_n} \prod_{j=2}^{n} \left\{ \sum_{s_j} \left( E_{s_i} + E_{\text{rad}_i} - E_{\text{rad}_j} + H_{\text{mol}} \right)^{-1} \left| s_j \right\rangle \left\langle s_j \right| \mathbf{m}_{i_{j-1}} \right\} \right| s_i \right\rangle. \qquad (8)$$

Note that the sum-notation $p[\{j\},n]$ signifies all possible combinations of the indices $(i_1,...,i_n)$.

### 3. Parametric susceptibilities

We now focus on parametric optical processes, where the matter and radiation field energies are conserved independently of each other (for example in second harmonic generation, each output photon is created at the expense of two input photons of half the frequency, while each optical centre returns to its original state, usually the ground state). Here, therefore, we have $s_f = s_i \equiv s_1$. Under these conditions the susceptibility tensor takes the form

$$O_{i_1...i_n}^{s_1 s_1} = \sum_{p[\{j\},n]} \left\langle s_1 \left| \mathbf{m}_{i_n} \prod_{j=2}^{n} \left( E_{s_i} + E_{\text{rad}_i} - E_{\text{rad}_j} + H_{\text{mol}} \right)^{-1} \mathbf{m}_{i_{j-1}} \right| s_1 \right\rangle \qquad (9)$$

In previous work based on a two-level approximation it has been shown that such tensors rigorously satisfy what we have loosely termed a 'mirror' rule [1]. This simply affirms that the molecular tensor for an elastic optical process in a pumped system is the negative of its ground state value, i.e. $O_{i_1...i_n}^{00} = -O_{i_1...i_n}^{uu}$. This result has been reported and proved explicitly both for the linear polarisability and the first hyperpolarisability, and it proves to remain valid even under near-resonance conditions that necessitate the inclusion of damping factors (see below). In general the mirror property can be expressed as $O_{i_1...i_n}^{00} + O_{i_1...i_n}^{uu} = 0$, physically signifying a bulk susceptibility that vanishes at the threshold of population inversion.

For many if not most systems, significant departures from such a rule can be anticipated, for two reasons: (i) the two-level approximation is seldom a good representation of a complex electronic structure – other levels should generally be considered; (ii) significant excited state populations are generally unsustainable except in association with laser action (where the dynamical ensemble exhibits harmonic resurgence features – which are nonetheless of considerable interest). In the following, we extend analysis to permit application to systems departing markedly from two-level behaviour, where the salient property sum is $\sum_{s_1} O_{i_1...i_n}^{s_1 s_1}$. It is not our purpose to identify exotic pumping schemes that might allow realisation of a vanishing ensemble susceptibility; our concern is with establishing the general validity of the new sum rule.

Before proceeding further, we note that it is possible to cast the generalised optical susceptibility tensor (9) in the form of an *operator* $\hat{O}_{i_n...i_1}$ in Hilbert space, whose diagonal elements are the results of equation (9);

$$\begin{aligned}\hat{O}_{i_n...i_1} &= \sum_{p[i_n,...,i_1]} \boldsymbol{m}_{i_n} \prod_{j=2}^{n}\left[\left(E_{s_i} + E_{\mathrm{rad}_i} - E_{\mathrm{rad}_j} + H_{\mathrm{mol}}\right)^{-1} \boldsymbol{m}_{i_{j-1}}\right] \\ &= \sum_{p[i_1,...,i_n]} \prod_{j=2}^{n}\left[\boldsymbol{m}_{i_j}\left(E_{s_i} + E_{\mathrm{rad}_i} - E_{\mathrm{rad}_j} + H_{\mathrm{mol}}\right)^{-1}\right]\boldsymbol{m}_{i_1} \\ &= \sum_{p[i_1,...,i_n]} \prod_{j=k+1}^{n}\left[\boldsymbol{m}_{i_j}\left(E_{s_i} + E_{\mathrm{rad}_i} - E_{\mathrm{rad}_j} + H_{\mathrm{mol}}\right)^{-1}\right]\boldsymbol{m}_{i_k} \prod_{j=2}^{k}\left[\left(E_{s_i} + E_{\mathrm{rad}_i} - E_{\mathrm{rad}_j} + H_{\mathrm{mol}}\right)^{-1}\boldsymbol{m}_{i_{j-1}}\right]\end{aligned}$$

(10)

This sum $\sum_{s_1} O_{i_1...i_n}^{s_1 s_1}$ is the Hilbert space trace of this operator; in the following sections we explicitly verify, for $n$ = 2, 3 and 4, that $\hat{O}_{i_1...i_n}$ is in fact traceless.

### 4. Two-photon process

The simplest case to consider is Rayleigh scattering. With a view to the more intricate cases to be examined subsequently, we introduce the molecular tensor as $O_{i_1 i_2}^{s_1 s_1}(\boldsymbol{h}_1 \hbar \boldsymbol{w}_1, \boldsymbol{h}_2 \hbar \boldsymbol{w}_2)$, where $\boldsymbol{h}_i = +1(-1)$ for an emitted (absorbed) photon. Although in this case the notation seems cumbersome, since $\boldsymbol{w}_1 = \boldsymbol{w}_2$ and $\boldsymbol{h}_1 = -\boldsymbol{h}_2$, its introduction will prove useful subsequently, as we entertain an increasing number of photon-matter interactions. To develop more concisely the Rayleigh case, we note that $-O_{i_1 i_2}^{s_1 s_1}(\boldsymbol{h}_1 \hbar \boldsymbol{w}_1, \boldsymbol{h}_2 \hbar \boldsymbol{w}_2) \equiv -O_{i_1 i_2}^{s_1 s_1}(-\boldsymbol{w}, \boldsymbol{w})$ signifies the linear electronic polarisability for a molecule in the state $s_1$, commonly written as $a_{i_1 i_2}^{s_1 s_1}(-w;w)$. Here the validity of the sum rule, when all molecular states are taken into account, is immediately evident from the following;

$$\sum_{s_1} O_{i_1 i_2}^{s_1 s_1}(h_1 \hbar w_1, h_2 \hbar w_2) = \sum_{s_1,s_2} \frac{\langle s_1|m_{i_1}|s_2\rangle\langle s_2|m_{i_2}|s_1\rangle}{E_{s_1 s_2} + h_2 \hbar w_2} + \sum_{s_1,s_2} \frac{\langle s_1|m_{i_2}|s_2\rangle\langle s_2|m_{i_1}|s_1\rangle}{E_{s_1 s_2} + h_1 \hbar w_1}$$

$$= \sum_{s_1,s_2} \frac{\langle s_1|m_{i_1}|s_2\rangle\langle s_2|m_{i_2}|s_1\rangle}{E_{s_1 s_2} + h_2 \hbar w_2} + \sum_{s_1,s_2} \frac{\langle s_2|m_{i_2}|s_1\rangle\langle s_1|m_{i_1}|s_2\rangle}{E_{s_2 s_1} + h_1 \hbar w_1}$$

$$= \sum_{s_1,s_2} \frac{\langle s_1|m_{i_1}|s_2\rangle\langle s_2|m_{i_2}|s_1\rangle}{(E_{s_1 s_2} + h_2 \hbar w_2)(E_{s_2 s_1} + h_1 \hbar w_1)} \left( \overbrace{E_{s_1 s_2} + E_{s_2 s_1}}^{=0} + \overbrace{h_1 \hbar w_1 + h_2 \hbar w_2}^{=0} \right)$$

(11)

where the permutation $p\{i_1, i_2\}$ has already been taken into account. Equation (11) proves that for this simple case the sum rule, $\sum_s a_{i_1 i_2}^{ss} = 0$, is valid. Elsewhere we have explicitly shown that the rule remains valid even when resonance damping is taken into account; each energy difference develops as $E_{s_1 s_2} \to \tilde{E}_{s_1 s_2} = E_{s_1 s_2} - i\hbar g_{s_1 s_2}$, where $g_{s_1 s_2} = g_{s_1} - g_{s_2}$ is the difference in damping factors for the states $s_1$ and $s_2$, and the imaginary terms cancel out in the sum exactly as in (11) above [2].

### 5.  Three-photon processes

It is not clear from the simple example given above that the sum rule is not a property uniquely associated with the simplicity of the Rayleigh process. We consider next a general three-photon parametric processes involves the photon set $\{h_1 w_1, h_2 w_2, h_3 w_3\}$. Processes in this category include second harmonic generation, sum-frequency generation and optical parametric down-conversion. In each case, as the process is elastic, members of the photon set satisfy the energy conservation condition $\sum_{i=1}^{3} h_i \hbar w_i = 0$. The susceptibility tensor is concisely expressible as $O_{i_1 i_2 i_3}^{s_1 s_1}{}^{(3)}(h_1 \hbar w_1, h_2 \hbar w_2, h_3 \hbar w_3) = b_{i_1 i_2 i_3}^{s_1 s_1}$ and its explicit expression, including all index permutations, is;

$$\sum_{s_1} b_{i_1 i_2 i_3}^{s_1 s_1} = \sum_{s_1 s_2 s_3} \frac{m_{i_3}^{s_1 s_3} m_{i_2}^{s_3 s_2} m_{i_1}^{s_2 s_1}}{\left(E_{s_1 s_3} + h_2 \hbar w_2 + h_1 \hbar w_1\right)\left(E_{s_1 s_2} + h_1 \hbar w_1\right)} + \frac{m_{i_3}^{s_1 s_3} m_{i_1}^{s_3 s_2} m_{i_2}^{s_2 s_1}}{\left(E_{s_1 s_3} + h_2 \hbar w_2 + h_1 \hbar w_1\right)\left(E_{s_1 s_2} + h_2 \hbar w_2\right)}$$

$$+ \frac{m_{i_1}^{s_1 s_3} m_{i_3}^{s_3 s_2} m_{i_2}^{s_2 s_1}}{\left(E_{s_1 s_3} + h_3 \hbar w_3 + h_2 \hbar w_2\right)\left(E_{s_1 s_2} + h_2 \hbar w_2\right)} + \frac{m_{i_1}^{s_1 s_3} m_{i_2}^{s_3 s_2} m_{i_3}^{s_2 s_1}}{\left(E_{s_1 s_3} + h_3 \hbar w_3 + h_2 \hbar w_2\right)\left(E_{s_1 s_2} + h_3 \hbar w_3\right)}$$

$$+ \frac{m_{i_2}^{s_1 s_3} m_{i_1}^{s_3 s_2} m_{i_3}^{s_2 s_1}}{\left(E_{s_1 s_3} + h_3 \hbar w_3 + h_1 \hbar w_1\right)\left(E_{s_1 s_2} + h_3 \hbar w_3\right)} + \frac{m_{i_2}^{s_1 s_3} m_{i_3}^{s_3 s_2} m_{i_1}^{s_2 s_1}}{\left(E_{s_1 s_3} + h_3 \hbar w_3 + h_1 \hbar w_1\right)\left(E_{s_1 s_2} + h_1 \hbar w_1\right)}$$

(12)

Analysis of the three-photon tensor serves as a helpful guide to the general principles for establishing the validity of the sum rule for any *n*-photon interaction tensor $\hat{O}_{i_1 \ldots i_n}$. The case will therefore be described in more detail below, with key statements for the three-photon case followed by equivalent statements for the general case $O_{i_1 \ldots i_n}$ given in square brackets.

From equation (12) it can be seen that there are 3! [*n*!] terms which can be grouped in two [(*n*-1)!] sets of three [*n*] terms each. The terms in each set are related to those in other sets by cyclic permutation. Explicitly, the first, third and fifth terms in (12) represent cyclic permutations of $(i_3, i_2, i_1)$, while the second, forth and sixth terms are cyclic permutations of $(i_1, i_2, i_3)$. Let us focus on the first set. Due to the fact that the labels of the molecular states $s_1$, $s_2$ and $s_3$ [$s_1$–$s_n$] *all* become dummy indices within sums over states, it is possible to rename them such that the third and fifth terms of (12) share the same numerator as the first;

$$\sum_{s_1 s_2 s_3} m_{i_3}^{s_1 s_3} m_{i_2}^{s_3 s_2} m_{i_1}^{s_2 s_1} \times \left\{ \frac{1}{\left(E_{s_1 s_3} + h_2 \hbar w_2 + h_1 \hbar w_1\right)\left(E_{s_1 s_2} + h_1 \hbar w_1\right)} \right.$$

$$+ \frac{1}{\left(E_{s_2 s_1} + h_3 \hbar w_3 + h_2 \hbar w_2\right)\left(E_{s_2 s_3} + h_2 \hbar w_2\right)} + \left. \frac{1}{\left(E_{s_3 s_2} + h_3 \hbar w_3 + h_1 \hbar w_1\right)\left(E_{s_3 s_1} + h_3 \hbar w_3\right)} \right\}.$$

(13)

When these three [*n*] terms are added using the energy conservation condition, $\sum_{i=1}^{3} h_i \hbar w_i = 0$, it follows that their sum vanishes;

$$\sum_{s_1 s_2 s_3} (-1) \, m_{i_3}^{s_1 s_3} \, m_{i_2}^{s_3 s_2} \, m_{i_1}^{s_2 s_1} \times \frac{\overbrace{E_{s_2 s_3} + E_{s_3 s_1} + E_{s_1 s_2}}^{=0} + \overbrace{h_3 \hbar w_3 + h_2 \hbar w_2 + h_1 \hbar w_1}^{=0}}{\left(E_{s_3 s_1} + h_3 \hbar w_3\right)\left(E_{s_2 s_3} + h_2 \hbar w_2\right)\left(E_{s_1 s_2} + h_1 \hbar w_1\right)} \ . \tag{14}$$

The second [second … $(n-1)^{th}$] set of terms in (12) cancels out in a similar way, proving that the sum rule is valid for any parametric three-photon [$n$-photon] interaction process. Again, we have elsewhere shown within the two-level approximation that the result remains valid when resonance damping is entertained [1].

### 6. Four-photon and higher order processes

The same procedure can be used to verify that the four-wave processes also satisfy the sum rule. In this case we have four photons satisfying the energy conservation condition $\sum_{i=1}^{4} h_i \hbar w_i = 0$ and the 24 terms in the third order susceptibility $O_{i_1 i_2 i_3 i_4}^{s_1 s_1 \, (4)}(h_1 \hbar w_1, h_2 \hbar w_2, h_3 \hbar w_3, h_4 \hbar w_4) = c_{i_1 i_2 i_3 i_4}^{s_1 s_1}$ can be separated into six sets of four terms each. It can be proved, again by renaming the dummy indices in the corresponding molecular operator, that the four terms comprising each set add to zero. As the number of photons involved in the optical process increases, the sum increasingly becomes algebraically cumbersome, and for this reason we refrain from presenting the explicit calculations. The sum rule remains generally valid;

$$\sum_s c_{i_1 i_2 i_3 i_4}^{ss} = 0 \tag{15}$$

The result applies to third harmonic generation, coherent anti-Stokes Raman scattering (CARS), laser-induced optical rotation, degenerate four-wave mixing (FWM) including phase conjugation, and indeed any other four-photon parametric process.

## 7. Discussion

A general proof of the susceptibility sum rule, valid for an arbitrary number of photon interactions, remains tantalisingly elusive, although the above explicit calculations offer important clues. The underlying principle merits further study, and we conclude by noting one novel but illuminating aspect, its fundamental relation to a thermodynamic limit. In a high-temperature, thermally equilibrated system, the fractional populations of all electronic states becomes equal, and for a parametric process the corresponding ensemble susceptibility is an equally weighted sum of the susceptibilities for all such states – which is thus zero. This directly relates to a principle that, at extremely high temperatures, the propensity for any material to mediate optical conversion becomes vanishingly small. Physically, this can be understood as an analogue of Curie's Law, and for the same underlying reason; as temperature increases, thermal motions increasingly interfere with the establishment of any polarisation field.

## Acknowledgments

This work is supported by a grant from the Engineering and Physical Sciences Research Council.